\documentstyle[amssymb,12pt,thmsa,sw20lart]{article}

\input{tcilatex}
\begin{document}

\title{Gradually Truncated Log-normal distribution - Size distribution of firms }
\author{Hari M. Gupta, Jos\'{e} R. Campanha \and Unesp - IGCE - Physics Dpto. \and %
Rio Claro - S\~{a}o Paulo - Brazil}
\maketitle

\begin{abstract}
Many natural and economical phenomena are described through power law or
log-normal distributions. In these cases, probability decreases very slowly
with step size compared to normal distribution. Thus it is essential to
cut-off these distributions for larger step size. Recently we introduce the
gradually truncated power law distribution to successfully describe
variation of financial, educational, physical and citation index. In the
present work, we introduce gradually truncated log-normal distribution in
which we gradually cut-off larger steps due to physical limitation of the
system. We applied this distribution successfully to size distribution of USA%
\'{}s manufactoring firms which is measured through their annual sell. The
physical limitation are due to limited market size or shortage of highly
competent executives. 
\end{abstract}

{\bf I. Introduction}

\bigskip

Many natural [1-16] and economical [17-23] phenomena are distributed through
power law [24,25] or log-normal [26] distribution. In log-normal
distribution, the logarithms of the larger steps are exponentially rare in
contrast to a normal distribution in which the larger steps themselves are
exponentially rare [27]. Thus the probability distribution decreases very
slowly with step size in compare to normal distribution as in power law
distribution. Recently we introduced gradually truncated power law
[23,28,29] in which we combine a statistical distribution factor and a
gradual cut-off after a certain step size which is due to the limited
physical capacity of the system to analyze the complex systems. In the
present paper we consider that larger steps of log-normal distribution
should also be gradually cut-off after certain step size like in power law
distribution. Thus, we propose a gradually truncated log-normal
distribution, which in line with gradually truncated power law [23] or
L\'{e}vy distribution [28] is described through:

\smallskip

\begin{equation}
P(\ln x)=\frac{1}{\sqrt{2\pi }\sigma }\exp (-\frac{(\ln x-\mu )^{2}}{2\sigma
^{2}})f(x)
\end{equation}

\smallskip

where $\mu $ and $\sigma $ are the mean and standard deviation of $\ln x.$
Further

\smallskip

\begin{equation}
f(x)=\left\{ 
\begin{array}{ccl}
1 & \mbox{if} & \left| x\right| \leqslant x_{c} \\ 
\exp \left\{ -\left( \frac{\left| x\right| -x_{c})}{k}\right) ^{\beta
}\right\} & \mbox{if} & \left| x\right| >x_{c}
\end{array}
\right.
\end{equation}

\smallskip

\noindent where x$_{C}$ is some cut-off value where physical limitations of
the system become important. $k$ and $\beta $ are constants of the system
for this factor. Though Central Limit theorem all distribution must approach
to normal distribution [27]. For log-normal distribution to approach normal
distribution for larger steps, we should have:

\smallskip

\begin{equation}
\beta =2
\end{equation}

\smallskip

Thus:

\smallskip

\begin{equation}
f(x)=\left\{ 
\begin{array}{ccl}
1 & \mbox{if} & \left| x\right| \leqslant x_{c} \\ 
\exp \left\{ -\left( \frac{\left| x\right| -x_{c})}{k}\right) ^{2}\right\} & %
\mbox{if} & \left| x\right| >x_{c}
\end{array}
\right.
\end{equation}

\smallskip

Generally, there is too little mass in the upper tail of the distribution.
However the statistical technique known as Zipf plot [30,31] which is a plot
of the log of the rank versus the log of the variable, is very useful to
discuss the distribution in the upper tail correctly

Let $(x_{1},x_{2},...,x_{N})$ be a set of $N$ observations on a random
variable x and supose that the observations are ordered from largest to
smallest so that the index $i$ is in the rank of $x_{i}$. The Zipf plot of
the sample is the graph of $ln$ $x_{i}$ against $ln$ $i$.

Thus $x_{i}$ may be estimated by the criterion [32]

\smallskip

\begin{equation}
\int_{x_{i}}^{\infty }N(x)dx=\int_{x_{i}}^{\infty }N.P(x)dx=i
\end{equation}

\smallskip

This specifies that there are $i$ number out of the ensemble $N$ which are
equal or more than $x_{i}$. From the dependence of $x_{i}$ on $i$ in a Zipf
plot, one can test whether it accord with a hypothesised form for $P(x)$. It
accentuates the upper tail of the distribution and therefore make it easier
to detect deviations in the upper tail from the theoretical prediction of a
particular distribution. Using Equation (1) and (5) and considering $P(\ln
x)d(\ln x)=P(x)dx$ we get:

\smallskip

\begin{equation}
i=\frac{N}{\sqrt{2\pi }\sigma }\int_{x_{i}}^{\infty }\frac{e^{-\frac{(\ln
x-\mu )^{2}}{2\sigma ^{2}}}}{x}f(x)dx
\end{equation}

\smallskip

or

\smallskip

\begin{equation}
\ln i=\ln (\frac{N}{\sqrt{2\pi }\sigma }\int_{x_{i}}^{\infty }\frac{e^{-%
\frac{(\ln x-\mu )^{2}}{2\sigma ^{2}}}}{x}f(x)dx)+C
\end{equation}

\smallskip

\noindent where C is a constant and is approximately equal to

\smallskip

\begin{equation}
C=\ln (\frac{N}{\sqrt{2\pi }\sigma })
\end{equation}

\smallskip

Now we discuss this distribution for size distribution of firms. This is
important because of their implications for the literature on the dynamics
of firm growth. Gibrat [26] showed that if the distribution of growth rates
is independent of firm size, the static distribution of firm size would
approach the log-normal. Hart and Prais [33] and Hall [34] found evidence
using British empirical data that the log-normal fits the distribution of
firm size reasonably well with some skewness to the rigth and that the
growth rates of firms were independence of initial size. Stanley et. al.
[35] using Zipf plot technique showed that although exist a agreement with
log distribution there is significant deviation for first 100 firms from
this distribution.

Here we take our empirical data from Stanley et. al. [35], which are annual
sell of 4071 Compustat firms in SIC codes 2000-3999 in year 1993. From
log-normal distribution plot of the sell of these firms, we consider $\mu
=17.76$ and $\sigma =2.72$ as also have been considered by them. We further
consider $x_{C}=8.10^{9}$ as deviation start from this point and $%
k=1.5.10^{11}.$ In Figure (1) we plot log rank versus log of sales. The
empirical result are in good agreement with present theory. For comparison
we also plot log-normal distribution. We especulate that this limiting
factor come from market capacity. If we consider log-normal distribution,
the sale of these firms must be of the order of $30.10^{12}$ dollars, about
three times the Gross national product of USA for this year, which is not
possible. Only first ten industries would have sell of the order of $%
20.10^{12}$ dollars i.e. about double of Gross National product. The total
sales of all industries through our model (gradually truncated log-normal)
comes out to be of the order of $10.10^{12}$ dollars, almost of the order of
Gross national product, which seems reasonable considering that: (i) Gross
National product includes all services and manufactured good. (ii) All
industries do not make final product. Many times final product of one
industry is initial product of other.

In conclusion sale of the very large firms depend not only on their inherent
potential to grow but also on available market, which however is not the
case for small industries, because enough market is available if they can
compete.We feel that like in power law distribution, the log-normal
distribution should also be gradually truncated due to physical limit of the
system.

{\bf Figure Captions}

\smallskip

\smallskip

\noindent {\bf Figure 1} Zipf plot double logarithmic plot of sales versus
rank for 4071 Compustat firms in SIC Codes 2000-3999 for year 1993 in USA.
....... represent empirical ,- - - - represent log-normal while -----
represents gradually truncated log-normal distribution

\smallskip \newpage


\smallskip

\newpage

\begin{thebibliography}{99}
\bibitem{}  A. Arneodo, J. F. Muzy and D. Sornette, Eur. Phys. J. {\bf B2},
227 (1998)

\bibitem{}  T. Lux and M. Marches, Nature {\bf 397}, 498 (1999)

\bibitem{}  C. K. Peng et al. Phys. Rev. Lett. {\bf 70}, 1343 (1993)

\bibitem{}  G. F. Zehende, P. M. C. de Oliveira and T. J. Penna, Phys. Rev.
E {\bf 57},3311 (1998)

\bibitem{}  J. B. Bassingthwaighte, L. S. Liebovitch and B. J. West, Fractal
Physiology (Oxford Univ. Press, New York 1994)

\bibitem{}  B. B. Mandelbrot, The Fractal Geometry of Nature (Freeman, New
York 1982)

\bibitem{}  U. Frish, M. F. Shlesinger and G. Zaslavasky, L\'{e}vy Flights
and Related Phenomena in Physics (Springer-Verlag, Berlin 1994)

\bibitem{}  T. H. Solomon, E. R. Weeks and H. L. Swinney, Phys. Rev. Lett. 
{\bf 71}, 3975 (1993)

\bibitem{}  M. Nelkin, Adv. Phys. {\bf 43}, 143 (1994)

\bibitem{}  A. Ott., J. P. Bouchard, D. Langevin and W. Urbach, Phys. Rev.
Lett. {\bf 65}, 2201 (1990)

\bibitem{}  Z. Olami, H. J. S. Feder and K. Christensen, Phys. Rev. Lett. 
{\bf 68}, 1244 (1992)

\bibitem{}  B. Chabaud et. al., Phys. Rev. Lett. {\bf 73}, 3227 (1994)

\bibitem{}  E. Weeks, J. Urbach and H. L. Swinney, Physica D {\bf 97}, 291
(1996)

\bibitem{}  T. H. Solomon, E. R. Weeks and H. L. Swinney, Physica D {\bf 76}%
, 70 (1994)

\bibitem{}  H. E. Hurst, Trans. Am. Soc. Civil Eng. {\bf 116}, 770 (1951)

\bibitem{}  C. Tsallis, Brazilian Journal of Physics {\bf 29}, 1 (1999)

\bibitem{}  Proceedings of First International Conference on High Frequency
Data in Finance (Olsen \& Associates, Zurich 1995)

\bibitem{}  R. N. Mantegna and H. E. Stanley, Nature {\bf 376}, 46 (1995)

\bibitem{}  G. Ghashghaie et. al., Nature {\bf 381}, 767 (1996)

\bibitem{}  E. F. Fama, Management Sci. {\bf 11}, 404 (1965)

\bibitem{}  A. L. Tucker, J. of Business \& Economic Statistics, {\bf 10},
73 (1992)

\bibitem{}  J. P. Bouchaud, M. Potters Theorie des risques financieres, Alea
Saclay (1997)

\bibitem{}  H. M. Gupta , J. R. Campanha and F. D. Prado, Int. J. Modern
Phys. C {\bf 11},1273 (2000)

\bibitem{}  P. L\'{e}vy, Th\'{e}orie de lAddition des Variables
Al\'{e}atories (Gauthier-Villars, Paris 1937)

\bibitem{}  G. Samorodnitsky and M. S. Taqqu, Stable Non-Gaussian Random
Processes: Stochastic Models with Infinite Variance (Chapman and Hall, New
York 1994)

\bibitem{}  R. Gibrat , Les in\'{e}galit\'{e}s economiques -Sirey Paris 1931

\bibitem{}  W. Feller, An Introduction to Probability Theory and Its
Applications (Wiley, New York 1971)

\bibitem{}  H. M. Gupta and J. R. Campanha, Physica A {\bf 268}, 231 (1999)

\bibitem{}  H. M. Gupta and J. R. Campanha, Physica A {\bf 275}, 531 (2000)

\bibitem{}  M. Gell-mann, The quark and the jaguar - W. H. Freeman, New York
-1994

\bibitem{}  G. K. Zipf, Human behavior and the principle of least effort -
Addison Wesley, Cambridge - 1949

\bibitem{}  J. Galambos, The Asymptotic Theory of extreme order Statistics
(J. Wiley \& Sons, New York 1978)

\bibitem{}  P. E. Hart and S. J. Prais, J. Royal Statistical Society Series
A, {\bf 119}, 150 (1956)

\bibitem{}  B. H. Hall, The Journal of Industrial Economics {\bf 35}, 583
(1987)

\bibitem{}  M. H. R. Stanley et. al. Economics Letters {\bf 49}, 453 (1995)

\pagebreak
\end{thebibliography}
\end{document}